\documentclass[12pt,preprint]{aastex}

\newcommand{\myemail}{zhekovs@colorado.edu}

\newcommand{\WR}{WR~147~}
\newcommand{\WRN}{WR~147N~}
\newcommand{\WRS}{WR~147S~}

\shorttitle{X-rays from \WR}
\shortauthors{Zhekov \& Park}

\begin{document}


\title{
{\it Chandra} Observations of \WR Reveal a Double X-ray Source
}


\author{Svetozar A. Zhekov\altaffilmark{1,3}
and Sangwook Park\altaffilmark{2} }


\altaffiltext{1}{JILA, University of Colorado, Boulder, CO
80309-0440, USA; \myemail}
\altaffiltext{2}{Department of Astronomy and Astrophysics,
Pennsylvania State University, 525 Davey Laboratory, University
Park, PA 16802, USA; park@astro.psu.edu}
\altaffiltext{3}{On leave from Space Research Institute, Sofia,
Bulgaria}


\begin{abstract}
We report the first results from deep X-ray observations of the
Wolf-Rayet binary system \WR with the {\it Chandra} HETG. Analysis of the
zeroth order data reveals that \WR is a {\it double} X-ray source.
The northern counterpart 
is likely associated with the colliding wind region, 
while the southern component is certainly identified with the WN
star in this massive binary.
The latter is the source of high energy X-rays
(including the Fe K$\alpha$ complex at 6.67~keV)
whose production mechanism is yet unclear.
For the first time, X-rays are observed {\it directly} from a WR star 
in a binary system.
\end{abstract}


\keywords{stars: individual (\WR) --- stars: Wolf-Rayet --- X-rays:
stars --- shock waves\\
~\vspace{0.1cm}
~\hspace{1in} {\bf To Appear in The Astrophysical Journal Letters}
}



\section{Introduction}
The Wolf-Rayet (WR) star \WR is a massive binary system that
is bright in X-rays and is a composite radio source.
Being the second closest WR star known allowed its emission to be 
spatially resolved in the radio, near infrared (NIR) and optical.
High-resolution radio observations showed that its southern component,
\WRS (the WN8 component in the binary), is a thermal source while its
northern counterpart, \WRN, is a non-thermal source
(\citealt{ab_86}; \citealt{mo_89}; \citealt{ch_92};
\citealt{con_96}; \citealt{wi_97}; \citealt{sk_99}).
High-resolution NIR data revealed two sources separated by 
$\approx 0\farcs64$ and \WRN was classified as a B0.5V star 
\citep{wi_97}.
An earlier spectral class, O8-O9 V-III, was
suggested from {\it Hubble Space Telescope} observations which 
spatially resolved the binary components \citep{nie_98}.
At a distance of $630\pm70$~pc to \WR \citep{ch_92},
the projected (or minimum) binary separation is $403\pm13$~au.
Since the binary components are massive stars, 
the interaction of their strong winds should form a colliding-stellar
wind (CSW) region (\citealt{pril_76}; \citealt{cherep_76}). 
Interestingly, from the analysis of radio and IR images, 
\citet{wi_97} concluded that the \WRN radio peak is
displaced by $0\farcs07$ southward from its IR counterpart.
This value is within the uncertainties of the data but 
if it is real, the non-thermal radio source can be associated with 
the CSW region located in vicinity of the OB companion.
The CSW scenario is supported by the analysis of the {\it Chandra} 
High Resolution Camera (HRC) image 
(although  with a very limited photon statistics and having no
spectral resolution) 
which concluded that the X-ray emission
is spatially extended and it peaks north of the WN8 star 
although a deeper X-ray image is needed to accurately determine 
the degree of spatial extension \citep{pitt_02}. 
It is worth noting that
contribution from the massive stars in the binary
cannot be excluded since these objects are X-ray sources themselves
(e.g. \citealt{gudel_naze_09} and the references therein).

X-ray observations are very important for studying CSWs.
They provide us with direct information about the physical conditions 
in the hot plasma behind strong shocks, which are likely the place
where non-thermal radio emission forms. The earlier X-ray
observations of \WR have revealed the presence of thermal
emission from high temperature plasma (\citealt{cai_85}; 
\citealt{sk_99}), but the CSW plasma characteristics were more 
tightly constrained only from the higher signal-to-noise
{\it XMM-Newton} data  \citep{sk_07}.
Skinner et al. reported a detection of the Fe K$\alpha$ complex at
6.67~keV and their analysis showed that the plasma temperature is
higher than the maximum temperature in the colliding-wind shock.
Consequently, \citet{zh_07} modeled the {\it XMM-Newton} spectra
successfully in the framework of the CSW scenario but only at the 
requirement  the stellar wind velocities were by a factor of  
$1.4 - 1.6$ higher than their currently accepted values. 

The fact that \WR is a wide binary at a relatively small distance
gives us an opportunity to examine the CSW phenomenon in more detail
that is to carry on spatially resolved X-ray studies. This
motivated our deep observations with the {\it Chandra} 
High Energy Transmission Gratings (HETG). In this {\it Letter} we 
report the first part of our analysis (based on the zeroth order data) 
which reveals a double X-ray source in \WR.

\section{Observations and Data Reduction}
\WR was observed with {\it Chandra} in the period Mar 28 - Apr 10,
2009. The observations in configuration  HETG-ACIS-S were carried out
in eight occasions
(\dataset[ADS/Sa.CXO\#obs/09941]{{\it Chandra} ObsIds: 9941,}
\dataset[ADS/Sa.CXO\#obs/09942]{ 9942,}
\dataset[ADS/Sa.CXO\#obs/10675]{10675,}
\dataset[ADS/Sa.CXO\#obs/10676]{10676,}
\dataset[ADS/Sa.CXO\#obs/10677]{10677,}
\dataset[ADS/Sa.CXO\#obs/10678]{10678,}
\dataset[ADS/Sa.CXO\#obs/10893]{10893 and}
\dataset[ADS/Sa.CXO\#obs/10897]{10897}) that provided 
7,300 zeroth order counts in a total
effective exposure of 286 ksec.
The CIAO 4.1.2 \footnote{Chandra Interactive Analysis of Observations 
(CIAO), http://cxc.harvard.edu/ciao/} data analysis software was used
through this study in combination with the {\it Chandra} calibration 
database CALDB v.4.1.3.
Following the CIAO Science Threads,
the original or combined event files were used for determining the
source coordinates, source variability and spectral extractions. 
By default, the pixel randomization is switched off in grating data.





\section{Analysis and Results}
\label{sec:results}
To have a full advantage of the total source counts in our
analysis of the zeroth order data, it is necessary to combine all data 
sets. As a prerequisite to this process, it is important  to estimate 
the accuracy of the source coordinates in the resultant total image.
This was done in two steps.

First, the CIAO command {\it wavdetect} was run on each of the eight
observations and an average position of the source was calculated.
The errors of the mean  were propagated from the individual errors 
derived from {\it wavdetect}.
Second, all the zeroth order data were merged eight times 
(CIAO command {\it merge\_all}) with a different reference image.
The source coordinates
were determined for each merged image by running {\it wavdetect}. 
The maximum difference in the source coordinates from one merged image
to another was $0\fs002$ and $0\farcs02$, correspondingly 
for the right ascension and declination.
The source positions are given in Table~\ref{tab:coor}.
We note that the center of the source
in the merged data is practically the same as the one
determined from averaging coordinates of the source centers derived for
each individual observation.
All this makes us confident that our analysis is not
influenced by the choice of the reference image.
The derived X-ray coordinates of \WR are
in good agreement (within the $1\sigma$ of the {\it Chandra}
positional 
accuracy \footnote{http://acs.harvard.edu/proposer/POG: \S 5.4.1}) 
with the ones determined for this object in the
optical ({\it HST} Guide Star Catalog) and in the radio 
(Table~\ref{tab:coor}).

\subsection{Images} 
\label{subsec:images}
The considerably better photon statistics (7,300 cts) of the 
total zeroth order HETG image compared to that of the previous  
{\it Chandra} HRC image ($\sim 148$~cts, \citealt{pitt_02})
gave us an opportunity to study in detail the spatial extent of the
X-ray emission from this object.
It also allowed us to carry on 
image analysis in different photon energy bands.

\begin{figure}[ht]
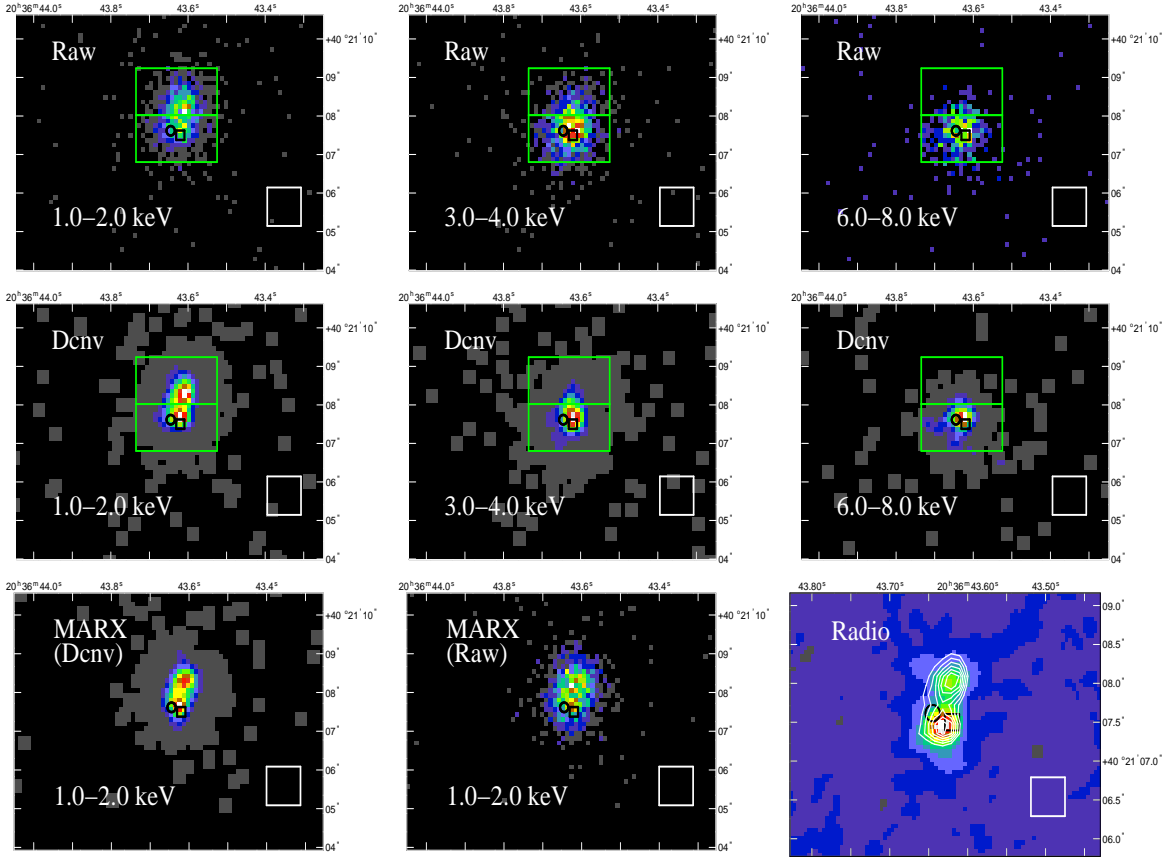

 \centering\includegraphics[width=2.in,height=1.5in,angle=-0]{f1a.eps}
 \centering\includegraphics[width=2.in,height=1.5in,angle=-0]{f1b.eps}
 \centering\includegraphics[width=2.in,height=1.5in,angle=-0]{f1c.eps}
 \centering\includegraphics[width=2.in,height=1.5in,angle=-0]{f1d.eps}
 \centering\includegraphics[width=2.in,height=1.5in,angle=-0]{f1e.eps}
 \centering\includegraphics[width=2.in,height=1.5in,angle=-0]{f1f.eps}
 \centering\includegraphics[width=2.in,height=1.5in,angle=-0]{f1g.eps}
 \centering\includegraphics[width=2.in,height=1.5in,angle=-0]{f1h.eps}
 \centering\includegraphics[width=2.in,height=1.5in]{f1i.eps}
\caption{
Examples of \WR X-ray images (pixel size of $0\farcs123$) 
in a linear scale by rows: 
(1) raw images;
(2) deconvolved ones; 
(3) images from MARX simulations and a radio image.
RA(J2000)  and Dec(J2000) are on horizontal and vertical axes,
respectively.
The optical position of \WR (HST GSC) is marked by a
circle and the position of the southern radio source \citep{con_99}
is marked by a square.
The scale of each image is illustrated by the
($1\arcsec\times1\arcsec$) square in its lower right  corner.
The 3.6-cm radio image of \WR is from a VLA observation on June
28, 1999.
The bright southern source is the WN8 star while the fainter emission
to its north is the non-thermal source (presumably arising in CSWs).
Overlaid are the
contours (linearly spaced by 0.1 of the maximum emission)
of the deconvolved X-ray image in the (1.0-2.0 keV) energy range.
The radio and X-ray images are in relative units normalized to their
corresponding maximum brightness. They were aligned that the
brightness peak of the southern X-ray source coincided with that of
the southern radio source.
The region boxes (in green) were used for estimating the relative
brightness of \WRN and \WRS (see Fig.~\ref{fig:spectra}). Their border
line defines how the two sources were separated for the spectral
extractions (Fig.~\ref{fig:extract}).
}
\label{fig:images}
\end{figure}

The relatively small spatial scale of the \WR emission in the radio
and optical 
($0\farcs57 - 0\farcs64$ which corresponds to
$\sim 1.2-1.3$ ACIS pixels)
prompted the use of subpixel images and 
image deconvolution techniques.
We applied a deconvolution procedure, based on a maximum likelihood
method (\citealt{rich_72}; \citealt{lucy_74}), that was successfully
used in the image analysis of SNR~1987A, an object with a small
spatial extent of $\sim1\farcs6$
(\citealt{Burr00}; \citealt{P02}, 2004, 2006; \citealt{judy09}).
The main result is that the morphology of the X-ray emission of \WR 
changes in different energy bands: its northern part is 
brightest in the (1.0-2.0 keV) range and gradually weakens at
higher energies. This likely indicates that a soft X-ray source is
located there while most of the hard energy photons originate in
the southern part of \WR.
This pattern is clearly seen from the raw subpixel X-ray images
as well as from the deconvolved ones (Fig.~\ref{fig:images}) 
and is also illustrated by the sources relative brightness
derived from direct photon counting (see lower right panel in 
Fig.~\ref{fig:spectra}).
It is thus conclusive that \WR is a {\it double}
X-ray source, consisting of a northern (\WRN) and a southern (\WRS)
part with different spectral characteristics. 
A separation of $\approx 0\farcs60$ between the positions of
maximum emission of \WRN and \WRS is directly measured from the
deconvolved (1.0 - 2.0 keV) image.

The existence of a spatial separation between \WRN and \WRS gets
additional support 
from measuring the \WR coordinates in different energy bands.
We ran the CIAO command {\it wavdetect} on the eight merged images.
We see that there is an offset of $\approx 0\farcs51$ between 
the source coordinates in the soft (1.0 - 2.0 keV) and hard 
(6.0 - 8.0 keV) energy bands (Table~\ref{tab:coor}). 
Interestingly, the source coordinates in the latter almost
perfectly coincide with the ones in the optical and radio. It is thus
conclusive that the X-ray hard-photon emitter (\WRS) is in fact the 
WR star itself (a WN8 object). Furthermore, a direct comparison of 
the \WR images in the radio\footnote{The radio data are from the 
NRAO Archive.
The National Radio Astronomy Observatory is a facility of the
National Science Foundation operated under cooperative agreement by
Associated Universities, Inc. }
and X-rays reveals a nice correspondence between the locations of
the radio  and X-ray \WRN sources (Fig.~\ref{fig:images}) which 
is suggestive that this is the CSW region in this binary system.

\subsection{Undispersed Spectra} 
\label{subsec:spectra}
\begin{figure}[ht]
 \centering\includegraphics[width=2.in,height=3.in,angle=-90]{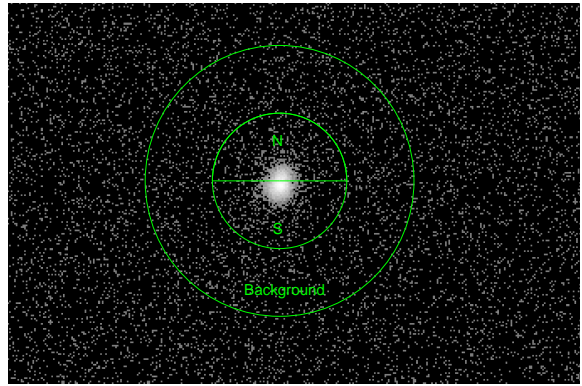}
\caption{
The extraction regions for the
total \WR,  \WRN and \WRS
spectra.
}
\label{fig:extract}
\end{figure}

Guided by the results from the image analysis, we extracted the X-ray 
spectra of \WR, \WRN and \WRS from the total (merged) zeroth order
data by using the CIAO script {\it specextract}
(for extraction regions see Fig.~\ref{fig:extract}). 
We fitted the spectra with
standard as well as custom models in version 11.3.2 of XSPEC
\citep{Arnaud96}.
We note that due to the small spatial separation between \WRN and \WRS
their spectra are not completely disentangled from each other. 
In our upcoming analysis of the first-order data, 
we will consider a more elaborated modeling of the undispersed spectra 
in conjunction with the dispersed ones. We adopted a simpler 
approach here which is nevertheless instructive for  the
emerging physical picture.

As seen from Fig.~\ref{fig:spectra}, the total spectrum of \WR is very
similar to that obtained with {\it XMM-Newton} \citep{sk_07}:
it is heavily absorbed, having most of its X-ray
counts at energies E~$ \geq 1$~keV. For consistency with the
previous studies, we adopted the same set of WN abundances (see 
\citealt{sk_07}; \citealt{zh_07}).
A simple 1T shock model ({\it vpshock} in XSPEC) gave a similar
quality of the fit as for the {\it XMM-Newton} spectra of \WR.
The X-ray absorption and the plasma temperature (with 90\%
confidence intervals in brackets) were
N$_H= 2.5 [2.3-2.9]\times10^{22}$~cm$^{-2}$,
kT~$=2.9 [2.3-3.2]$~keV, and the derived abundances fell within the
90\% confidence intervals of the {\it XMM-Newton} spectral fits
\citep{sk_07}. 
The total observed flux was F$_X(0.5-10~keV) =
1.28\times10^{-12}$~erg~s$^{-1}$,
about 15\% smaller than that from the {\it XMM-Newton}
spectra.
A better constraint on this difference would be possible after 
completing upcoming analysis of the first-order HETG data.

Exploring the physical picture that identifies \WRN with the
CSW region in the binary system, 
the X-ray spectra of \WRN and \WRS were fitted simultaneously and
they shared the same WN abundances set. 
The latter is justified by that the shocked WN wind dominates the
X-ray emission of the CSW region (\citealt{zh_07}; for the use of
discrete temperature models see \S~5.2 therein).
The 1T shock model with individual postshock
temperatures for  \WRN and \WRS and a common X-ray absorption gave a 
poor quality of the fit (reduced $\chi^2 \approx 1.5$). 
An acceptable fit was obtained if 
the spectra had individual
X-ray absorption (see Fig.~\ref{fig:spectra} and
Table~\ref{tab:fits}).
It should be emphasized that in both sources the plasma temperature
well exceeds 1 keV.

As in the image analysis, we see that the northern source is
softer than its southern counterpart. Also, the latter (the WN8 star) 
is the place where the Fe K$\alpha$ complex at 6.67~keV comes from
(see the inset in Fig.~\ref{fig:spectra}).
\WRS is subject to an excess X-ray absorption compared to \WRN.
Applying the \citet{go_75} conversion
(NH $= 2.22\times10^{21}$A$_V$~cm$^{-2}$), the absorption towards 
\WRN almost perfectly corresponds to the optical extinction of \WR
(A$_v = 11.6$~mag, A$_V =$~A$_v/1.11$; \citealt{vdh_01}).

We thus note that the results from image and spectral analysis seem
consistent with a physical picture where \WRN resides in the CSW
region of the binary, while the X-ray source \WRS is
likely located deeper in the WR wind.

\subsection {MARX Simulations} 
\label{subsec:marx}
We used version 4.4 of the
 MARX\footnote{See http://space.mit.edu/CXC/MARX/} software to
simulate the observational situation with \WR.
We ran a 286-ksec HETG exposure for a X-ray source
composed of two point sources with spectral characteristics
corresponding to those of \WRN and \WRS (Table~\ref{tab:fits}).
The sources were located at the positions of maximum emission for the
northern and southern sources as in the (1.0-2.0 keV) deconvolved
image of \WR. 
We note a nice correspondence between the simulated and observed data
(Fig.~\ref{fig:images}).

All this makes us more confident about the results from our analysis,
that illustrate the superior capabilities of the {\it Chandra} 
observatory even when working on their very edge.






\section{Discussion}  
Analysis of the zeroth order HETG data showed that {\it Chandra}
resolved \WR into a double X-ray source. 
Its two counterparts, \WRN and \WRS, are most likely 
identified correspondingly 
with the CSW region and the WN8 star in this wide binary system.
We recall that the maximum temperature in the CSW region is determined
by the terminal wind velocities in the binary and for \WR
their currently accepted values could not provide the plasma
temperature deduced from the X-ray spectra 
\citep{sk_07}. This is why,
successful CSW models of \WR were possible only for wind velocities
being a factor of $1.4-1.6$ higher \citep{zh_07}. 
We note that all this is based on analysis 
of the unresolved {\it XMM-Newton} data which assumed
that the CSW
region is responsible for the {\it total} X-ray emission from \WR. 
The high resolution {\it Chandra} data resolve this discrepancy
by revealing two sources with different spectral characteristics.

\begin{figure}[ht]
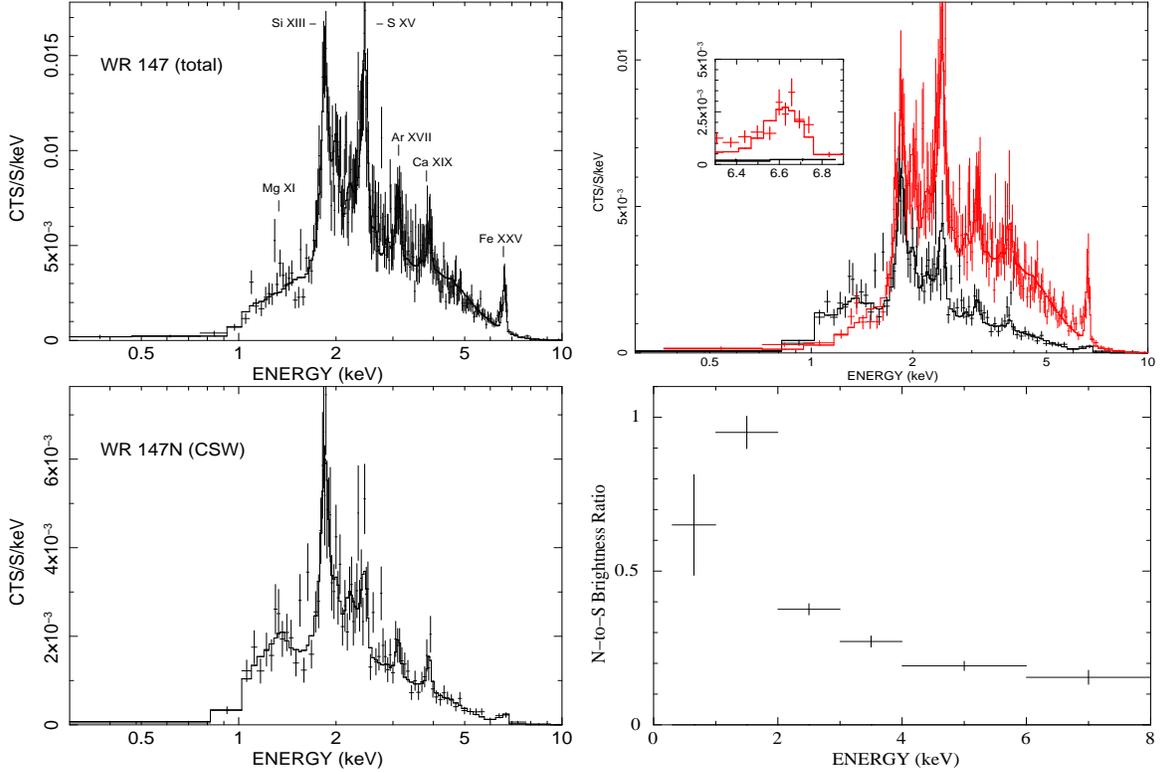

 \centering\includegraphics[width=2.in,height=3.in,angle=-90]{f3a.eps}
 \centering\includegraphics[width=2.in,height=3.in,angle=-90]{f3b.eps}
 \centering\includegraphics[width=2.in,height=3.in,angle=-90]{f3c.eps}
 \centering\includegraphics[width=2.in,height=3.in,angle=-90]{f3d.eps}
\caption{
WR 147 background-subtracted spectra (rebinned to have a minimum
of 20 counts per bin).
{\it Upper left panel:} 
The total spectrum overlaid with the best-fit
1T shock model ($\chi^2/dof = 269/230$).
Prominent emission lines are marked.
{\it Upper right panel:} 
the \WRN and \WRS (in red) spectra and the
1T shock models with individual X-ray absorption.
The inset shows the Fe K$_{\alpha}$ complex at 6.67 keV.
{\it Lower left panel:} 
the X-ray spectrum of \WRN overlaid
with the CSW model ($\chi^2/dof = 98/80$).
%
{\it Lower right panel:} 
The \WRN-to-\WRS relative brightness defined 
in different energy ranges (marked with horizontal bars)  
by counting the number of photons in the two region boxes as 
shown in Fig.~\ref{fig:images}.
}
\label{fig:spectra}
\end{figure}

We used the CSW model with the nominal stellar wind parameters 
(V$_{WR} = 950$~km~s$^{-1}$, $\dot{M}_{WR} = 4\times 10^{-5}$ 
M$_{\odot}$~yr$^{-1}$; 
V$_{O} = 1600$~km~s$^{-1}$, $\dot{M}_{O} = 6.6\times 10^{-7}$
M$_{\odot}$~yr$^{-1}$;
$[\dot{M}_{O} V_{O} / \dot{M}_{WR} V_{WR}] = 0.028$;
for CSW model details see \citealt{zh_07}) to fit the X-ray spectrum 
of \WRN. As seen from Fig.~\ref{fig:spectra},
the shape of the observed spectrum is perfectly matched, 
that is there is NO temperature discrepancy between the  model and 
observations any more.
But, the model overestimates the total luminosity of \WRN
by a factor of $\sim 16$. We note that this discrepancy is in general
trackable. The CSW X-ray luminosity scales with the mass loss rate
$\dot{M}$, wind velocity $v$ and binary separation as
$L_X \propto \dot{M}^2 v^{-3} D^{-1}$ (\citealt{luo_90};
\citealt{mzh_93}). If the mass-loss rates are factor of $\sim2$ or
even more lower than assumed because of their intrinsic uncertainties
or because the winds are clumped (e.g. \citealt{cr_07}), this can 
account for most of the luminosity mismatch, and the unknown
orbital inclination (larger binary separation) would account for the
difference. All this will be explored in detail in our upcoming
combined modeling of the first-order spectra and the undisperesed
ones.
%

While this gives more confidence that \WRN resides in the CSW region
of the binary system, an identification of \WRN with the OB star in
the system cannot be completely ruled out.
Our analysis of the X-ray spectral lines (from the first-order
HETG spectra)
will be crucial in this respect. 
For example, if suppressed forbidden line in the He-like triplets
is found (an indication of high density or strong UV emission), 
it will favor the OB star
origin for \WRN.
Unfortunately, the luminosity of the OB companion is poorly
constrained \citep{lepine_01} which prevents estimating its 
contribution
to the X-ray emission of \WRN based on the L$_X$ - L$_{bol}$ relation. 
But, the OB stars are in general soft X-ray sources 
with plasma temperatures less than 1 keV
(see \S 4.1.2 and \S 4.3 in the review paper of
\citealt{gudel_naze_09}).
Given the high X-ray absorption to \WR (\S~\ref{subsec:spectra}),
we thus anticipate that this contribution will not be appreciable
unless the OB star in \WR is a rare hot magnetic object (see \S 4.6 in
\citealt{gudel_naze_09}).

On the other hand, the fact that the WN8 star is a 
hard-energy X-ray source (\WRS) is indeed puzzling.
Skinner et al. (2002a,b) have discussed in some detail possible
mechanisms for X-ray production in WN stars when 
the high-energy tail was established with certainty in the X-ray
emission from presumably single WR stars: WR 6 and WR 110.
These mechanisms include: instability-driven wind shocks (IDWS); 
magnetically confined wind shocks (MCWS); wind 
accretion shocks; colliding wind shocks (including the case of
stellar wind shocking onto a close companion); 
non-thermal X-ray emission.
We note that none of these mechanisms finds solid observational
support for the moment and 
each of them has its own limitations and caveats.
For example, the IDWS should be considerably soft X-ray
emitters while the MCWS might be able to provide the high plasma
temperature observed in \WRS. But for the massive stellar winds in
WRs, this requires a relatively strong global magnetic field in the
star itself. At present, such fields have not been reported for
Wolf-Rayet stars.
Presence of a strong magnetic field may also suggest flare 
activity but no short-term variability is detected in any of the eight HETG 
observations (the CIAO tool {\it glvary} returns a variability index 
of zero). The same was found in earlier X-ray observations
of \WR (\citealt{sk_99}; \citealt{sk_07}).
Presence of a compact (or a normal star) companion is intriguing but 
this may suggest that we have quite a rare opportunity to observe a triple 
system that initially consisted of at least two or even three 
massive stars and in the latter case the most massive one has already 
exploded.

Finally, although the mechanism responsible for the high-energy
X-ray production is unclear it may not be unique to \WR.
Apart from earlier detections (WR 6  and WR 110), 
\citet{sk_09} report a hard-energy 
tail in the X-ray spectra of several more, presumably single, WN stars.
It is worth noting that all these WNs have a subtype different
from \WR while a star of its subtype, WR 40 (a WN8 star), was not
detected with {\it XMM-Newton} \citep{gosset_05}.
Hopefully, future observations will facilitate solving the mystery of 
this puzzling phenomenon: how common the high-energy tail is for 
single WR stars and how it correlates with their subtype.
But, \WR is unique in the sense that for the first time
we detect X-rays {\it directly} from a WR star in a binary system.
Moreover, the upcoming analysis of the first-order HETG spectra will 
give us a chance to measure line parameters 
(widths, centroids, fluxes) in the X-ray spectrum {\it intrinsic} to a 
WR star (\WRS), thus, to study its X-ray plasma characteristics in
detail.




\acknowledgments
This work was supported by NASA through Chandra grant GO9-0013A to
the University of Colorado at Boulder, and through grant G09-0013B to
the Pennsylvania State University.



{\it Facilities:} \facility{CXO (HETG, ACIS)}.

\clearpage



\clearpage

\begin{deluxetable}{lllc}
\tablecaption{\WR Positions: X-ray, optical, radio \label{tab:coor}}
\tablewidth{0pt}
\tablehead{
\colhead{} &
\colhead{$\alpha_{2000}$} & \colhead{$\delta_{2000}$} &
\colhead{Uncertainty}\\
\colhead{} &
\colhead{$20^h\, 36^m$} & \colhead{$+40\arcdeg\, 21\arcmin$} &
\colhead{($\alpha, \delta$)}
}
\startdata
Average        & $43\fs636$ & $ 7\farcs69$  & $\pm 0\farcs02$  \\
Merged         & $43\fs636$ & $ 7\farcs71$  & $\pm 0\farcs01$  \\
(1.0 - 2.0 keV)& $43\fs631$ & $ 7\farcs95$  & $\pm 0\farcs03$  \\
(6.0 - 8.0 keV)& $43\fs648$ & $ 7\farcs48$  & $\pm 0\farcs02$  \\
HST GSC        & $43\fs64$  & $ 7\farcs62$  & $\pm 0\farcs32$  \\
Radio$^{a}$    & $43\fs62$  & $ 7\farcs5$   & $\pm 0\farcs05$  \\
Radio$^{b}$    & $43\fs64$  & $ 7\farcs5$   & $\pm 0\farcs01$  \\
\enddata
\tablecomments{The source position by rows:
(1) the mean from the eight {\it Chandra} observations;
(2) from the merged images;
(3, 4) from `filtered' merged images;
(5) optical;
(6, 7) radio.
}
\tablenotetext{a,b}{
The radio coordinates (\citealt{con_99}; \citealt{wi_97})
are for \WRS (the WN8 star)
converted from B1950 into J2000 with the HEASARC coordinate converter
(http://heasarc.gsfc.nasa.gov/cgi-bin/Tools/convcoord/convcoord.pl).
}
\end{deluxetable}

\begin{deluxetable}{lcc}
\tablecaption{1T Shock Model Results 
\label{tab:fits}}
\tablewidth{0pt}
\tablehead{
\colhead{} & \colhead{\WRN}  & \colhead{\WRS} 
}
\startdata
$\chi^2$/dof  & \multicolumn{2}{c}{309/256}  \\
N$_H$ ({\footnotesize 10$^{22}$ cm$^{-2}$}) &  
        2.28 [2.08 - 2.57] & 3.83 [3.51 - 4.20] \\
kT (\footnotesize{keV})    &  1.78 [1.52 - 1.98] & 2.36 [2.12 - 2.56] \\
Ne   & \multicolumn{2}{c}{19.7 [0.0 - 76.2]}  \\
Mg   & \multicolumn{2}{c}{3.3 [0.5 - 6.7] }   \\
Si   & \multicolumn{2}{c}{5.1 [3.9 - 7.2]}   \\
S~   & \multicolumn{2}{c}{6.7 [5.5 - 7.8]}   \\
Ar   & \multicolumn{2}{c}{8.4 [5.8 - 9.8]}  \\
Ca   & \multicolumn{2}{c}{8.3 [4.3 - 12.4]}  \\
Fe   & \multicolumn{2}{c}{10.0 [7.7 - 11.8]}  \\
F$_X$$^a$ ({\footnotesize $10^{-12}$ ergs cm$^{-2}$ s$^{-1}$})  
            & 0.324 (4.4) & 0.921 (12.3)  \\
\enddata
\tablecomments{
Brackets enclose 90\% confidence intervals.
All abundances
are with respect to their solar values (\citealt{an_89}).
The fixed in the fit abundances are: 
H$ = 1$, He$ = 25.6$, C$ = 0.9$, N$ = 140$, O$ = 0.9$, and Ni$ = 1$
(for details see \citealt{sk_07}; \citealt{zh_07})
}

\tablenotetext{a}{
The observed X-ray flux (0.5 - 10 keV) followed in parentheses
by the unabsorbed value. 
}
\end{deluxetable}




\end{document}